\documentclass[fleqn,twoside]{article}
\usepackage{espcrc2}

\usepackage{graphicx}
\newcommand{\AmS}{{\protect\the\textfont2
  A\kern-.1667em\lower.5ex\hbox{M}\kern-.125emS}}

\newcommand{\dslash}[2]{{{#1}\hspace{-5pt}{/}}_{#2}}
\newcommand{\Bsl}{{B\hspace{-5pt}{/}}}
\newcommand{\ndop}{{\mathcal O}_{R/L;L}^{\Bsl}}
\newcommand{\msbar}{\overline{\mbox{MS}}}

\title{
\vspace{-3.6cm}
\begin{flushright}
{\normalsize
hep-lat/0210008\\
RBRC-252\\
}
\end{flushright}
\vspace{1.5cm}
Nucleon Decay Matrix Elements for Domain-Wall Fermions}

\author{Yasumichi Aoki\address{RIKEN BNL Research Center, Brookhaven
National Laboratory, Upton, NY 11973, USA} [RBC collaboration]
\thanks{We thank RIKEN, Brookhaven National Laboratory and the U.S.\
Department of Energy for providing the facilities essential for the
completion of this work.}}

\begin{document}

\begin{abstract}
We report on the nucleon decay matrix elements with 
domain-wall fermions in quenched approximation. 
Results from direct and indirect method are compared with a focus
on the process of a proton decaying to a pion and a lepton.
We discuss the renormalization necessary for the matching 
to the continuum theory. Preliminary results for 
the renormalized chiral lagrangian parameters are presented.
%\vspace{-12pt}
\end{abstract}

\maketitle

\section{INTRODUCTION}

Nucleon decay is one of the most important aspect that 
any (SUSY) GUT model has.
In low-energy effective theories it is represented by 
dimension six operators
made of one lepton field and three quark fields.
In the decay of $p\to \pi^0+e^+$, for example, 
we need to calculate hadronic matrix element of the
% SU(3)gauge-invariant 
three quark 
operator $\ndop =\epsilon^{ijk}(u^{iT} CP_{R/L}d^j)P_{L}u^k$ with initial
proton and final pion states, 
where $C$ is the charge conjugation matrix. $P_{R/L}$ represents the
right or left projection matrix, respectively.

Lattice calculation of the nucleon decay matrix elements historically
started with the indirect calculation \cite{NDSummary}.
First, one estimates the overlap of the operator to the proton state,
$\alpha$ and $\beta$,
\begin{eqnarray}
 \alpha P_L u_p  & \equiv &
  \langle 0 | {\mathcal O}_{R;L}^{\dslash{B}{}} | p;\vec{k}=\vec{0}\rangle,\\
 \beta P_L u_p & \equiv &
 \langle 0 | {\mathcal O}_{L;L}^{\dslash{B}{}} | p;\vec{k}=\vec{0}\rangle,
\end{eqnarray}
with $u_p$ being the proton spinor.
Then, with the help of chiral perturbation theory, one can calculate the
nucleon decay matrix elements.
The first attempt to calculate the nucleon decay matrix element
directly handling the three point function is done by Gavela et.~al
\cite{Gavela:1989cp}, 
showing significant deviation from the indirect 
calculations. Recently JLQCD \cite{Aoki:1999tw} 
pointed  out the incompleteness of the calculation \cite{Gavela:1989cp}. 
The matrix element has a tensor structure,
\begin{equation}
 \langle \pi;\vec{p} | \ndop | p;\vec{k}\rangle = 
  P_{L} [ W_0 - i \dslash{q}{} W_q] u_p,
\end{equation}
where $q=k-p$. The relevant form factor
$W_0$ is what we need since the term proportional to $\dslash{q}{}$ 
vanishes after multiplying the lepton spinor.
$W_q$ is called the irrelevant form factor.
JLQCD uses quenched Wilson gauge configurations and Wilson
fermion at parameters summarized in Table~\ref{tab:param} (c).
The three quark operators are renormalized with one-loop 
perturbation theory. Now the difference of direct and indirect
calculations is not huge, but, still $30$ to $40\%$ for most cases.

In this study we try to calculate nucleon decay matrix elements with
domain-wall fermions in the quenched approximation.
Using DBW2 gauge action
% \cite{Takaishi:1996xj,deForcrand:1999bi}
makes chiral symmetry breaking especially 
small compared to the other actions \cite{Orginos:2001xa,Aoki:2001dx}.
With this good chiral symmetry one expects the good 
property to calculate hadronic matrix elements: 1) preventing operator
mixing with different chiral structure, and 2) good scaling, even down to
$1/a\simeq 1$ GeV region 
\cite{Blum:2000kn,Aoki:2001dx}. 
We restrict ourselves to the case of degenerate quark mass
in the meson. In this case, 
one can calculate $p\to\pi^0+l^+$ and $p\to\pi^++\bar{\nu}$ decay
amplitudes. The latter is obtained by multiplying the former
by $\sqrt{2}$  under the exact SU(2) symmetry of  $u$ and $d$ quarks.
%
%\vspace{-12pt}

\section{PROCEDURE AND RESULTS}

\begin{figure}[htb]
   \begin{center}
    \includegraphics[width=7cm]{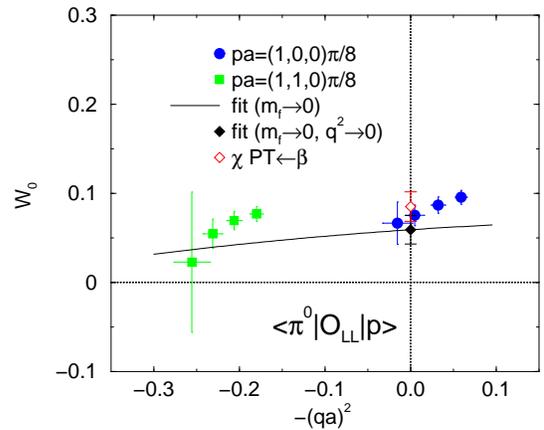}
   \end{center}
\vspace{-32pt}
 \caption{$W_0$ for $p\to (\pi^0,l^+)$ decay with ${\mathcal O}_{L;L}$ as
 a function of invariant mass squared of lepton. }
\label{fig:res}
\vspace{-12pt}
\end{figure}

% *(50 conf) L_s=12, m_{res}=1.16(5) 10^{-3}, $m_\rho(m_f=-m_{res})=0.625(24)
% (100 conf) L_s=16, m_{res}=0.57(3) 10^{-3}, $m_\rho(m_f=-m_{res})=0.589(20)
%
\begin{table*}[t]
 \caption{Lattice parameters and corresponding inverse lattice spacings
 with $\rho$ mass input. Values of $r_0$ are listed for 
 another estimate of the scale. We use the parameter set (a) for the
 nucleon decay. $\dag$ and $\ddag$ represent unpublished data with 50
 and 100 configurations respectively.
}
 \label{tab:param}
\begin{center}
 \begin{tabular}{ccccccccccc}
  \hline
  & fermion & size ($V\times T$) & $L_s$ &
  $M_5$ & $1/a_\rho$[GeV] & ref. & gauge & $6/g^2$ & $r_0/a$ & ref. \\ 
  \hline
  (a) & {\bf DWF} & {\bf $16^3\times 32$} & {\bf 12} & {\bf 1.8} &
    1.23(5) & $\dag$ & {\bf DBW2} & {\bf 0.87} & & \\ 
  (b) & DWF & $16^3\times 32$ & 16 & 1.8 & 1.31(4) & \cite{Aoki:2001dx} &
    DBW2 & 0.87 &
    \multicolumn{1}{c}{\raisebox{1.5ex}[0pt]{3.58(4)}} &
    \multicolumn{1}{c}{\raisebox{1.5ex}[0pt]{$\ddag$}} \\
  \hline
  (c) & Wilson & $28^3\times 48\times 80$ & - & - &
    2.30(4) & \cite{Aoki:1999tw} & Wilson & 6.0 & & \\ 
  (d) & DWF & $16^3\times 32$ & 16 & 1.8 & 1.92(4) & \cite{Blum:2000kn} &
    Wilson & 6.0 &
    \multicolumn{1}{c}{\raisebox{1.5ex}[0pt]{5.368}} & 
    \multicolumn{1}{c}{\raisebox{1.5ex}[0pt]{\cite{Guagnelli:1998ud}}} \\
  \hline
 \end{tabular}
\end{center}
 \vspace{-8pt}
\end{table*}
The parameters of our simulation are shown in Table~\ref{tab:param} (a).
Our quark propagator is obtained by averaging two propagators
calculated using periodic and anti-periodic boundary conditions. Thus, the 
effective temporal extent is $64$. 100 independent gauge
configurations are analyzed for the three and two point functions 
to calculate the ratio \cite{Aoki:1999tw},
\begin{equation}
 R(t)=\frac{\langle J_{\pi}(t_1) \ndop (t)
  \bar{J_{p}}(t_0)\rangle}
  {\langle J_{\pi}(t_1) J_{\pi}^{\dagger}(t)\rangle
  \langle J_{p}(t) \bar{J_{p}}(t_0)} \sqrt{Z_\pi Z_p}.
  \label{eq:ratio3pt}
\end{equation}
In the three point function proton and pion interpolating fields are
located at $t_0=6$ and $t_1=24$ respectively.
Momentum $\pm\vec{p}$ with $\vec{p}a=(1,0,0)\pi/8$ or $(1,1,0)\pi/8$ is
injected to the pion and the operator in the  
three point function, as well as in the pion two point function 
in the denominator.
% The quark source of $J_{p}(t_0)$ is opimized to the nuceon: 
% the source vector is uniformly distributed
% in a $8^3$ box in $t_0$ time slice, where Coulomb-gauge fixing
% is done.
$\sqrt{Z_\pi}$ and $\sqrt{Z_p}$ 
are overlap of $J_{\pi}$ and $J_{p}$ 
% with the point source quark propagator 
to the corresponding pion and proton states, which 
is estimated from the fit of two point functions.
We take four different quark masses $m_fa=0.02$, $0.04$, $0.06$, $0.08$,
among which the strange mass is located between the last two.
% We need to extract 
% $W_0$ which always mixes with $W_q$ as we always have to project 
% out the positive parity sub space.
%
% Figure \ref{fig:ratio} shows the ratio (\ref{eq:ratio3pt}) with
% appropreate projection and trace after subtraction of the irrelevant
% contiribution for a parameter set. We make a fit always to the range 
% $10\le t\le 15$ for $W_0$. Fig.~\ref{fig:res} is the result for 
% the $\langle \pi^0|{\mathcal O}_L;L|p\rangle$ as a function of $(qa)^2$.
%
After appropriate projection, trace and the subtraction of the
irrelevant contribution,
$W_0$ is calculated with a fit to plateau of the ratio.
The results with bare operator ${\mathcal O}_{L;L}^{\Bsl}$ are shown in
Figure \ref{fig:res}. 
$W_0$ can only depend on $q^2$ and on $m_f$ which can enter through
the masses of proton and pion.
We assume a form of the fitting function of $W_0=c_0 + c_1 q^2 + c_2
(q^2)^2 + c_3 m_f$. In our precision the physical kinematics are
approximated as $m_f=0$ and $-q^2=0$. The filled diamond in the figure
% represents the result of it.
represents the result extrapolated to this point.
We also calculate the chiral lagrangian parameters $\alpha$ and $\beta$
extrapolated to $m_f=0$. 
%
%The chiral lagrangian parameters $\alpha$ and $\beta$ are calculated 
%form the ratio of the two point functions,
%%
%\begin{equation}
% R^{\alpha/\beta}(t) = \frac{{\mathcal O_{L; R/L}^\Bsl(t)}
%  \bar{J_{p}}(t_1)}
%  {\langle J_{p}(t) \bar{J_{p}}(t_1)} \sqrt{Z_p}.
%  \label{eq:ratio2pt}
%\end{equation}
%%
%The value obtained with the fit to plateau of this ratio 
%are extrapolated to $m_f=0$ with a fitting function linear in $m_f$.
%
By the tree-level chiral perturbation theory \cite{Claudson:1982gh},
the relevant part of the matrix element is obtained,
% as,
%
\begin{equation}
 \langle \pi^0 | {\mathcal O}_{L; L}^\Bsl | p \rangle_{\mbox{rel}}
  \simeq \beta (1+D+F) P_L u_p / \sqrt{2}f,
  \label{eq:chL_LL}
\end{equation} %\\
% D+F &= &1.27.
%
where $D=0.47$ and $F=0.80$ from the experiment \cite{Hsueh:1988ar},
and $f$ is $f_\pi=0.131$ GeV.
The open diamond in Fig.~\ref{fig:res} represents this
indirect estimation.
%
%$\alpha$ and $\beta$ calculated by the ratio (\ref{eq:ratio2pt})
%at finite $m_f$ are extraporated to $m_f=0$ with a fitting 
%function linear in $m_f$. The open diamond in Fig.~\ref{fig:res}
%is obtained with $\beta$ substituted into eq.~(\ref{eq:chL_LL}). 
%
There is no significant difference between direct and indirect results
in the current statistics. However, $40\%$ excess for the central value
from indirect method is quite similar to the JLQCD result\cite{Aoki:1999tw}.
Similar of result is obtained also for 
$\langle \pi^0 | {\mathcal O}_{R; L}^\Bsl | p \rangle$.

The above results need be renormalized for the value in the $\msbar$ 
scheme in continuum theory.
We expect 
these operators are multiplicatively renormalized as the chiral symmetry
breaking is small
 ($m_{res}\sim 1.5$ MeV).
 The renormalization factor for DBW2 gauge action 
is available for one-loop perturbation theory \cite{Aoki:2002iq} with 
$L_s$ treated as infinity.
In Table \ref{tab:Z} we list the values of renormalization factors for
the nucleon decay by perturbation theory with two different
mean field (MF) improvement schemes. The MF factor $u$ is
either $P^{1/4}$ or $1/8K_c$, where $P$ stands for plaquette,
$K_c$ the critical hopping parameter of the four dimensional Wilson
fermion. We also list the renormalization factor for the axial
vector current, for which we know the non-perturbative value at
the parameter point of (b) in Table~\ref{tab:param} with the method
using the conserved current \cite{Blum:2000kn}.
\begin{table}[t]
 \caption{Renormalization factors for the axial vector current and
 the nucleon decay three-quark operator in $\msbar$ with NDR.
 Scale $\mu=1/a$ is chosen for nucleon decay.}
 \label{tab:Z}
\begin{center}
 \begin{tabular}{c|cc|c}
  \hline
  & \multicolumn{2}{c|}{perturbative\cite{Aoki:2002iq}} & non \\
  \multicolumn{1}{c|}{\raisebox{1.5ex}[0pt]{operator}} & plaq & $K_c$ &
  perturbative\\
  \hline
  $\bar{q}\gamma_5\gamma_\mu t^a q$  & 0.976 & 0.769 & 0.7776 (5)\\
  $\mathcal O_{L; R/L}^\Bsl$         & 1.009 & 0.713 & - \\
  \hline
 \end{tabular}
\end{center}
 \vspace{-24pt}
\end{table}
For the axial vector current renormalization, the mean field improvement
with $K_c$ is in better agreement with the non-perturbative one, 
while the other is 30\% larger. This difference is considered
as the systematic error of the perturbative estimate.
The difference arises mainly from the quark wave function
renormalization and the MF factor. Thus, this problem does not apply
to quantities defined by a particular ratio in which those factors
cancel out. $B_K$ is one of those cases.
\begin{table}[b]
\vspace{-14pt}
 \caption{Chiral lagrangian parameter renormalized at $\mu=1/a$
 in $\msbar$ in NDR. Scale from $\rho$ mass is used for results in
 [$\mbox{GeV}^3$]. Quoted errors are statistical only.}
 \label{tab:amp}
% \hspace{-12pt}
 \begin{tabular}{c|cc|cc}
   \hline
   & \multicolumn{2}{c|}{$[\mbox{GeV}^3]\times 10^{-2}$} & & \\
   & $|\alpha|$ & $|\beta|$ & $|\alpha| r_0^3$ & $|\beta| r_0^3$ \\
   \hline
   This work & 0.6(1) & 0.7(1) & 0.15(3) & 0.18(4) \\
   JLQCD\cite{Aoki:1999tw} & 1.5(1) & 1.4(1) & 0.19(2) & 0.18(2)\\
   \hline
 \end{tabular}
\end{table}

In Table \ref{tab:amp} we list the result of $\alpha$ and $\beta$ of
ours and by JLQCD. The operators are renormalized at $\mu=1/a$ with $\msbar$
for both results. Renormalization factor with $K_c$ is employed 
for our calculation.
Our values of $\alpha$ and $\beta$ in [$\mbox{GeV}^3$] obtained
by setting the scale with the $\rho$ mass input, are quite different
from those with Wilson fermion by JLQCD. 
The origin of the discrepancy could be, 
1) discretization error, 2) systematic error of the MF perturbation,
3) difference of the renormalization scale.
2) is about 30\% for DWF, 3) is almost negligible.
1) is expected to be most severe as it is seen in the difference
of the scale from $\rho$ mass for Wilson and DWF with Wilson gauge
background at same $1/g^2$ (see (c) and (d) in Table \ref{tab:param}).
Indeed, if we set the scale from different quantity $r_0$, which
eliminates order $a$ error in the scale for Wilson fermion, the values
are consistent. 

%This difference is quite big even if
%the systematic error of the MF perturbation for DWF, and the
%difference of the renormalization scale, whose effect is expected
%to be much smaller, are taken into account.
%The difference could be caused by the order $a$ scaling violation
%of the $\rho$ mass for Wilson fermion (compare $a_\rho$ of (c) and (d) 
%in Table \ref{tab:param}). If one uses gluonic scale $r_0$, 
%which has no order $a$ error, the results are consistent.

\section{SUMMARY}
We calculated proton decay matrix elements with DBW2 gauge action and
DWF.
We did not find a significant difference in values obtained
by direct and indirect methods within our current numerical precision.
Further simulation with
non-degenerate quark masses will reduce the error of direct calculation
and make the extraction of nucleon to kaon decay amplitudes possible.
Within the current
systematic error of the perturbation theory the chiral lagrangian
parameters for the nucleon decay calculated with DWF are consistent with 
those obtained with Wilson fermion by JLQCD if the scale is set by $r_0$.
More stringent comparison will be done for those values
after non-perturbative renormalization is performed.

%\bibliography{proc}

\begin{thebibliography}{9}
\bibitem{NDSummary}
{\it For summary and references, see} \cite{Aoki:1999tw}.

\bibitem{Gavela:1989cp}
M.B. Gavela et~al.,
\newblock Nucl. Phys. B312 (1989) 269.

\bibitem{Aoki:1999tw}
JLQCD, S. Aoki et~al.,
\newblock Phys. Rev. D62 (2000) 014506, hep-lat/9911026.

\bibitem{Orginos:2001xa}
RBC, K. Orginos,
\newblock Nucl. Phys. Proc. Suppl. 106 (2002) 721, hep-lat/0110074.

\bibitem{Aoki:2001dx}
RBC, Y. Aoki,
\newblock Nucl. Phys. Proc. Suppl. 106 (2002) 245, hep-lat/0110143.

\bibitem{Blum:2000kn}
T. Blum et~al.,
\newblock (2000), hep-lat/0007038.

\bibitem{Guagnelli:1998ud}
ALPHA, M. Guagnelli, R. Sommer and H. Wittig,
\newblock Nucl. Phys. B535 (1998) 389, hep-lat/9806005.

\bibitem{Claudson:1982gh}
M. Claudson, M.B. Wise and L.J. Hall,
\newblock Nucl. Phys. B195 (1982) 297.

\bibitem{Hsueh:1988ar}
S.Y. Hsueh et~al.,
\newblock Phys. Rev. D38 (1988) 2056.

\bibitem{Aoki:2002iq}
S. Aoki et~al.,
\newblock (2002), hep-lat/0206013.

\end{thebibliography}
%\bibliographystyle{h-elsevier}

\end{document}